\documentclass[a4paper,11pt]{article}
\usepackage{amsmath,amssymb,mathrsfs,amsthm}

\theoremstyle{plain}
\newtheorem{thm}{Theorem} 
 
\newtheorem{prop}{Proposition}[section]
\newtheorem{lemma}[prop]{Lemma}

\theoremstyle{definition} 
  
\newtheorem{asm}{Assumption}

\theoremstyle{remark}
\newtheorem{remark}[prop]{Remark}

\newcommand{\RR}{\mathbb{R}}
\newcommand{\CC}{\mathbb{C}}
\newcommand{\ZZ}{\mathbb{Z}}
\newcommand{\TT}{\mathbb{T}}
\newcommand{\Cinf}{C^{\infty}}
\newcommand{\ve}{\varepsilon}
\newcommand{\vp}{\varphi}
\newcommand{\supp}{\mathop{\mathrm{supp}}}
\newcommand{\RE}{\mathrm{Re}\,}
\newcommand{\IM}{\mathrm{Im}\,}
\newcommand{\bcdot}{\boldsymbol{\cdot}}
\newcommand{\Sch}{Schr\"{o}dinger }

\begin{document}

\title{Continuum limit of resonances for discrete \Sch operators}
\author{Kentaro Kameoka\footnote{Research Organization of Science and Technology, Ritsumeikan University, 
1-1-1, Nojihigashi, Kusatsu-shi, Shiga, Japan 525-8577. E-mail: kkameoka@fc.ritsumei.ac.jp}
 \and Shu Nakamura\footnote{Department of Mathematics, Faculty of Sciences, Gakushuin University, 1-5-1, Mejiro, Toshima, Tokyo, Japan 171-8588, Email: shu.nakamura@gakushuin.ac.jp}}

\date{}

\maketitle

\begin{abstract}
We consider complex resonances for discrete and continuous Schr\"odinger operators, 
and we show that the resonances of discrete models converge to resonances of continuous models 
in the continuum limit. The potential is supposed to be a sum of an exterior dilation analytic 
function and an exponentially decaying function, which may have local singularities. 
The proof employs a generalization of the norm resolvent convergence of discrete Schr\"odinger 
operators by Nakamura and Tadano (2021), combined with the complex distortion method in the Fourier space. 
Our results confirm that the complex resonances can be approximately computed using discrete Schr\"odinger 
operators. We also give a recipe for the construction of approximate discrete operators for Schr\"odinger operators 
with singular potentials. 
\end{abstract}

\section{Introduction}

In this paper, we characterize resonances of 
\[
H=-\Delta+V(x)\,\,\,\,\mathrm{on}\,\,\, L^2(\RR^d)
\]
by the continuum limit of resonances of its discretization: 
\[
H_h=T_h+V_h(x)\,\,\,\,\mathrm{on}\,\,\, \ell^2(h \ZZ^d), 
\]
where $h>0$ is a small parameter and 
\[
T_hu(x)=-h^{-2}\sum_{|x-y|=h}(u(y)-u(x)).
\] 
The discretization $V_h: h\ZZ^d\to \RR$ is constructed from $V: \RR^d\to \RR$, which is described later. 
In the first part of this paper, we suppose $V$ is continuous, and in this case $V_h$ is simply defined as 
the restriction  of $V$ to $h\ZZ^d$. 
In the second part, we consider a singular $V$,  and then $V_h$ is defined as the restriction of a regularization of $V$ to $h\ZZ^d$.
Our definition of resonances employs the complex distortion in the Fourier space. 

In this paper, we denote the Fourier transform by
\[
\mathcal{F} u(\xi)=\hat{u}(\xi)=(2\pi)^{-d/2}\int_{\RR^d} u(x)e^{-ix\cdot \xi}dx.
\]  
We also write $\langle x \rangle=(1+|x|^2)^{1/2}$ for $x\in\RR^d$, etc.

\subsection{Continuous potentials}\label{subsec-1.1}
We first consider a continuous $V$ and assume the following. 

\begin{asm}\label{asm-1}
The potential $V$ is of the form $V=V_1+V_2$ with the following properties. 
\begin{enumerate}
\renewcommand{\labelenumi}{(\arabic{enumi})\,}
 \item $V_1\in \Cinf(\RR^d; \RR)$ and it is analytic in $\{x+iy\in \CC^d|\, |x|>R_0, |y|<c_0|x|\}$ for some $R_0>0$ and 
$0<c_0\le 1$ with the bound $|V_1(x+iy)|\lesssim |x|^{-\mu}$ for some $\mu>0$ in that region. 
 \item There exists $\delta_0>0$ such that the Fourier transform $\hat{V_2}$ is analytic in 
$\{\xi+i\eta \in \CC^d|\, |\eta|<\delta_0\}$ and 
\[
\sup_{|\eta|<\delta'_0}\int\limits_{\RR^d}|\hat{V_2}(\xi+i\eta)|d\xi<\infty  
\]
for any $0<\delta'_0<\delta_0$. 
\end{enumerate}
\end{asm}

\begin{remark}
Assumption~\ref{asm-1},~(2) implies that $V_2$ is continuous and 
$|V_2(x)|\lesssim e^{-\delta'_0 |x|}$ for any $0<\delta'_0<\delta_0$ by the Paley-Wiener theory. 
Conversely, if there exists $\delta_0>0$ and $s>d/2$ such that  $e^{\delta_0\langle x \rangle}V_2\in H^s(\RR^d; \RR^d)$, 
then $V_2$ satisfies Assumption~\ref{asm-1},~(2).  
\end{remark}

We set 
\[
V_h=V|_{h\ZZ^d},\quad \text{i.e.},\quad V_h(x)=V(x) \ \text{ for }x\in h\ZZ^d, 
\]
which is well-defined since $V$ is continuous. 
We set 
\[
\Omega(\delta_0, c_0)=\{z\in \CC|\, \IM z>-2\delta_0 |\RE z|^{1/2}, -2 \arctan c_0 <\arg z<\pi/2\}
\] 
for $\delta_0,c_0>0$. 
We denote the set of resonances of $H$ and $H_h$ by $\mathrm{Res}(H)$ and $\mathrm{Res}(H_h)$.  
Resonances of $H$ are the poles of meromorphic continuation of 
$(u_1, (z-H)^{-1}u_2)$, $\IM z>0$, for $u_1, u_2\in L^2_{\mathrm{comp}}(\RR^d)$. 
Similarly, resonances of $H_h$ are the poles of meromorphic continuation of 
$(u_1, (z-H_h)^{-1}u_2)$, $\IM z>0$, for compactly supported $u_1, u_2\in \ell^2(h\ZZ^d)$. 
Multiplicities of resonances are defined suitably. 
Then resonances of $H$ are defined in $\Omega(\delta_0, c_0)$, 
and for any $\Omega \Subset \Omega(\delta_0, c_0)$ resonances of $H_h$ 
are defined in $\Omega$ for sufficiently small $h>0$ (see~Section~2 for details on the definition of resonances). 

\begin{thm}\label{thm-1}
Under Assumption~\ref{asm-1}, 
\[
\mathrm{Res}(H)=\lim_{h\to 0}\mathrm{Res}(H_h)
\]
in $\Omega(\delta_0, c_0)$ including multiplicities. 
\end{thm}
The statement means that for any $z\in \Omega(\delta_0, c_0)$ and 
small $\gamma>0$, there exists $h_0>0$ such that for $0<h<h_0$, 
the number of resonances of $H_h$ in $\{\zeta\in \CC|\, |\zeta-z|<\gamma\}$ counted with their 
multiplicities is equal to the multiplicity of $z$ as a resonance of $H$, which is zero if $z \not\in \mathrm{Res}(H)$. 
This in particular implies the Hausdorff distance between $\mathrm{Res}(H_h)$ and $\mathrm{Res}(H)$ tends to zero 
locally when $h\to 0$.

\subsection{Singular potentials}\label{subsec-1.2}
We now consider a singular $V$. 
We first define its regularization with a smooth 
rapidly decreasing function $\phi\in \mathscr{S}(\RR^d; \RR)$.   
We suppose $\int_{\RR^d} \phi(x) dx=1$ and we suppose 
$\hat{\phi}$ is analytic and rapidly decreasing in 
$\{\xi+i\eta\in \CC^d|\,|\eta|<C\}$ for any $C>0$.  
For instance, we may take $\phi\in C_0^\infty(\RR^d;\RR)$, or $\phi(x)=(2\pi)^{-d/2}e^{-x^2/2}$. 
We set $\phi_h(x)=h^{-d}\phi(x/h)$.  
We define $V_h=(\phi_h\ast V)|_{h\ZZ^d}$, 
and we set $H_h=T_h+V_h$ on $\ell^2(h\ZZ^d)$ in this subsection. 

\begin{asm}\label{asm-2}
The potential $V$ is of the form $V=V_1+V_2$ with the following properties. 
\begin{enumerate}
\renewcommand{\labelenumi}{(\arabic{enumi})\,}
\item  $V_1\in \Cinf(\RR^d; \RR)$ and it is analytic in $\{x+iy\in \CC^d|\, |x|>R_0, |y|<c_0|x|\}$ for some $R_0>0$ and 
$0<c_0\le 1$ with the bound $|V(x+iy)|\lesssim |x|^{-\mu}$ for some $\mu>0$ in that region. 
 \item There exists $\delta_0>0$ such that the Fourier transform $\hat{V_2}$ is analytic in 
$\{\xi+i\eta \in \CC^d|\, |\eta|<\delta_0\}$ and 
\[
\sup_{|\eta|<\delta'_0}\int\limits_{\RR^d}|\hat{V_2}(\xi+i\eta)|^p d\xi<\infty  
\]
for any $0<\delta'_0<\delta_0$. 
Here $p=2$ for $d=1, 2, 3$ and $1<p<d/(d-2)$ for $d\ge 4$. 
\end{enumerate}
\end{asm}

\begin{remark}
When $d=1, 2, 3$, Assumption~\ref{asm-2},~(2) is equivalent to the condition that 
$e^{\delta'_0 |x|}V_2\in L^2(\RR^d; \RR^d)$ for any $0<\delta'_0<\delta_0$ as is well-known.  
When $d\ge 4$, if there exist $\delta_0>0$ and  $s\ge 0$ with $s>d/2-2$ such that 
$e^{\delta_0\langle x \rangle}V_2\in H^s(\RR^d; \RR^d)$, then $V_2$ satisfies Assumption~\ref{asm-2},~(2). 
For instance, the Coulomb potential with several fixed nuclei on $\RR^3$ satisfies Assumption~\ref{asm-2} for
any $0<c_0<1$ and any $\delta_0>0$.  
\end{remark}

\begin{thm}\label{thm-2}
Under Assumption~\ref{asm-2}, 
\[
\mathrm{Res}(H)=\lim_{h\to 0}\mathrm{Res}(H_h)
\]
in $\Omega(\delta_0, c_0)$ including multiplicities. 
\end{thm}

We also note that our proof without the complex distortion shows 
the continuum limit for discrete eigenvalues for singular potentials. 

\begin{thm}\label{thm-3} 
Assume that $\hat{V}\in L^1+L^2$ for $d=1, 2, 3$ and $\hat{V}\in L^1+L^p$ with $1<p<d/(d-2)$ for $d\ge 4$. 
Take any $\phi\in \mathscr{S}(\RR^d;\RR)$ with $\int \phi(x)dx=1$ 
and set $V_h=(\phi_h\ast V)|_{h\ZZ^d}$. 
Then there exists a family of isometries $I_h: \ell^2(h\ZZ^d)\to L^2(\RR^d)$, $0<h<1$, such that 
\[
\lim_{h\to 0}\|I_h(H_h-z)^{-1}I_h^*-(H-z)^{-1}\|_{L^2\to L^2}=0
\]
for any $z\in \CC\setminus\sigma(H)$. 
In particular, $\lim \sigma_d(H_h)=\sigma_d(H)$ including multiplicities. 
\end{thm}

\begin{remark}
The above assumption is satisfied if $V=V_1+V_2$ and there exist real numbers 
$s_j\ge 0$ with $d/2-2<s_j<d/2$ such that $V_j\in |D|^{-s_j}L^2$ for $j=1, 2$, 
where $|D|^{-s_j}L^2$ is the homogeneous Sobolev space of order $s_j$. 
Example includes the Coulomb potential with several fixed nuclei on $\RR^3$.  
Such results are analogous to the results in Nakamura-Tadano \cite{NT}, 
but the potentials are always supposed to be continuous in the previous literature. 
\end{remark}

\begin{remark}
Straightforward modifications of the proofs show that Theorem~\ref{thm-2} and Theorem~\ref{thm-3} hold true 
even if we set $\phi_h(x)=h^{-\alpha d}\phi(x/h^{\alpha})$ with $0<\alpha<1$. 
This setting might be relevant for certain discrete approximation scheme. 
\end{remark}

Resonances correspond to quasi-steady states and are related to scattering theory. 
Mathematical study of resonances is a vast field of research (see, for example, textbooks: Cycon, Froese, Kirsch, Simon \cite{CFKS} and Dyatlov, Zworski \cite{DZ}). 
Resonances for discrete \Sch operators have been studied by a few authors especially for the one-dimensional case 
(see, for instance, Klopp \cite{Kl}). 

The complex distortion in the Fourier space was studied by Cycon \cite{C} and Sigal \cite{S} for 
radially symmetric dilation analytic potentials and sufficiently smooth exponentially decaying potentials. 
Nakamura \cite{N}  extended the method to not necessarily radially symmetric potentials. 
The dilation analytic part is assumed to be globally dilation analytic in \cite{N} while we include 
exterior dilation analytic potentials such as the Coulomb potential in this paper. 
 
Continuum limit for discrete \Sch operators have been studied recently by several authors. 
See, for instance, Nakamura, Tadao \cite{NT},  Isozaki, Jensen \cite{IJ} and Mikami, Nakamura, Tadano \cite{MNT}. 
In particular, the generalized norm resolvent convergence is proved in \cite{NT}, which implies the 
convergence of discrete eigenvalues (cf. Theorem~\ref{thm-3}). 
In this paper, we generally follow the argument of \cite{NT}, though the details are significantly more 
involved because of the complex distortion and the lack of continuity. 
We also refer Cornean, Garde, Jensen \cite{C-G-J, C-G-J-3}, Miranda, Parra \cite{MirPar}, 
Exner, Nakamura, Tadano \cite{ExNaTad} for generalizations of the results of \cite{NT}. 
Continuum limit of Dirac operators are discussed in Cornean, Garde, Jensen \cite{C-G-J-2} and 
Nakamura \cite{N2024}. 
We may investigate the continuum limit of resonances for Dirac operators and 
other models studied in the above papers using our method. 

This paper is organized as follows. 
In Section~2, we recall the Fourier distortion for continuous \Sch operators and 
discuss its analogue for discrete \Sch operators. 
In Section~3, we prove Theorem~\ref{thm-1}. 
In Section~4, we prove Theorem~\ref{thm-2} and Theorem~\ref{thm-3}. 


\section{Complex distortion in the Fourier space}
\subsection{Fourier distortion for continuous \Sch operators}\label{subsec-1}
In this subsection, we recall the complex distortion in the Fourier space for 
continuous \Sch operators (see~\cite{N}). 
Complex distortion for the discrete \Sch operator is defined analogously which is discussed in the next subsection. It is then used in the proof of the continuum limit of resonances.  
Our Assumption~\ref{asm-2} is more general than that in \cite{N} and includes exterior dilation analytic potentials 
such as the Coulomb potential. 

We write $T=-\Delta$ and $\widetilde{T}(\xi)=\xi^2$. 
Then 
\[
\widetilde{H}=\mathcal{F}H\mathcal{F}^{-1}=\widetilde{T}(\xi)+\widetilde{V},
\] 
where $\widetilde{V}$ is the convolution operator by $(2\pi)^{-d/2}\hat{V}$. 
We fix a vector field $v\in \Cinf(\RR^d; \RR^d)$. 
We assume $v$ is compactly supported, partly for simplicity, 
but also for later applications. 
We set 
\[
\Phi_{\theta}(\xi)=\xi+\theta v(\xi), \quad 
U_{\theta}f(\xi)=J_{\theta}(\xi)^{1/2}f(\Phi_{\theta}(\xi)),\quad \xi\in\RR^d, 
\]
where $J_{\theta}(\xi)$ is the Jacobian: $\det\Phi_{\theta}'(\xi)$. 
We define 
\[
\widetilde{H}_{\theta}=U_{\theta} \widetilde{H} U_{\theta}^{-1}
=\widetilde{T}(\Phi_{\theta}(\xi))+\widetilde{V}_{\theta}, 
\]
where $\widetilde{V}_{\theta}$ has the integral kernel 
\[
\widetilde{V}_{\theta}(\xi, \eta)=(2\pi)^{-d/2}J_{\theta}(\xi)^{1/2}\hat{V}
(\Phi_{\theta}(\xi)-\Phi_{\theta}(\eta))J_{\theta}(\eta)^{1/2}. 
\] 
We also set 
\[
H_{\theta}=\mathcal{F}^{-1}\widetilde{H}_{\theta}\mathcal{F},\quad  
T_{\theta}=\mathcal{F}^{-1}\widetilde{T}(\Phi_{\theta}(\xi))\mathcal{F}, \quad
V_{\theta}=\mathcal{F}^{-1}\widetilde{V}_{\theta}\mathcal{F}. 
\] 
We note that by Assumption~\ref{asm-2}, $\widetilde{V}_{\theta}(\xi, \eta)$ is analytic with respect to $\theta$. 

\begin{lemma}\label{lemma-kernel}
Under Assumption~\ref{asm-1} or Assumption~\ref{asm-2},
\[
|\hat{V}(\Phi_{\theta}(\xi)-\Phi_{\theta}(\eta))|\le \Psi(\xi-\eta),   
\]
where $\Psi \in L^1(\RR^d)$ under Assumption~\ref{asm-1}, 
and $\Psi \in L^1(\RR^d)+L^p(\RR^d)$ under Assumption~\ref{asm-2}. 
Here $\theta$ belongs to 
\[
\bigl\{\theta\in \CC\bigm| \mathrm{Lip}(v)(c_0|\RE\theta|+|\IM\theta|)<c_0, 
\|v\|_{\infty}|\IM \theta|<\delta_0\bigr\}, 
\]
where $\mathrm{Lip}(v)$ is the Lipschitz constant of $v$
\end{lemma}
\begin{proof}
We first consider $V_1$ in Assumption~\ref{asm-1} or Assumption~\ref{asm-2}.  
We see that $\hat{V_1}$ has an analytic continuation to 
$\{\xi+i\eta\in \CC^d|\, |\eta|<c_0|\xi|\}$ by the contour deformation of 
the integral in the definition of the Fourier transform. 
When $\xi$ ranges over a small conic neighborhood of $\xi_0$ and $|\eta|<c'_0|\xi|$,  
the contour is of the form 
$\bigl\{x-ic''_0(1-\chi(x))|x|\frac{\xi_0}{|\xi_0|} \bigm| x\in \RR^d\bigr\}$, where $0<c'_0<c''_0<c_0$ and 
$\chi\in \Cinf_c(\RR^d)$ is equal to $1$ on a large compact set. 
By a change of variables, we may assume $\xi_0=(1, 0, \dots, 0)$ and the contour deformation is justified by 
the Cauchy's theorem with respect to $x_1$. 
Then the smoothness of $V_1$ implies the decay of $\hat{V_1}$ 
and the decay of $V_1$ implies the bound $|\hat{V_1}(\xi+i\eta)|\lesssim |\xi|^{-d+\mu}$ near the origin if $0<\mu<d$. 
Note that $\Phi_{\theta}(\xi)-\Phi_{\theta}(\eta)$ belongs to 
\[
\bigl\{\zeta\in \CC^d\bigm| |\IM \zeta|+c'_0|\RE \zeta-(\xi-\eta)|\le 
(c'_0\mid\RE \theta|+|\IM \theta|)\mathrm{Lip}(v)|\xi-\eta|\bigr\}, 
\] 
which is included in $\{\zeta\in \CC^d|\,|\IM \zeta|<c'_0|\RE \zeta|\}$ if 
\[
\mathrm{Lip}(v)(c'_0|\RE\theta|+|\IM\theta|)<c'_0<c_0.
\] 
Thus 
\[
|\hat{V}(\Phi_{\theta}(\xi)-\Phi_{\theta}(\eta))|\lesssim |\xi-\eta|^{-d+\mu}\quad \text{for }|\xi-\eta|<1.
\] 
To estimate $\hat{V}(\Phi_{\theta}(\xi)-\Phi_{\theta}(\eta))$ for $|\xi-\eta|\gg 1$,
we use the Minkowski's inequality, $\|v\|_{\infty}|\IM \theta|<\delta_0$ 
and the fact that $\hat{V}$ is harmonic, which completes the proof. 
\end{proof}

\begin{prop}\label{prop-1}
Under Assumption~\ref{asm-2}, $H_{\theta}$ with $D(H_{\theta})=H^2(\RR^d)$ 
is an analytic family closed operators of type(A) and $H_{\theta}^*=H_{\bar{\theta}}$.  
The spectrum of $H_{\theta}$ is discrete in the unbounded component of 
$\CC\setminus \{\Phi_{\theta}(\xi)^2|\, \xi\in \RR^d\}$. 
Here $\theta$ ranges over $\{\theta\in \CC|\mathrm{Lip}(v)(c_0|\RE\theta|+|\IM\theta|)<c_0, 
\|v\|_{\infty}|\IM \theta|<\delta_0\}$. 
\end{prop}
\begin{proof}
It is enough to prove 
$\widetilde{V}_{\theta}(\xi^2+1)^{-1}$ is compact and analytic with respect to $\theta$. 
Then $H_{\theta}^*=H_{\bar{\theta}}$ follows from the Kato-Rellich theorem and 
the discreteness of the spectrum follows from the Fredholm theory.  

We take $q>1$ with $p^{-1}+q^{-1}=1$ and $q_1>1$ with $1/2+q^{-1}=q_1^{-1}$. 
Then $q=2$ for $d=1, 2, 3$ and $q>d/2$ for $d\ge 4$. 
Since $(\xi^2+1)^{-1}\in L^{\infty}(\RR^d)\cap L^q(\RR^d)$, H\"{o}lder's inequality implies that 
$(\xi^2+1)^{-1}: L^2(\RR^d)\to L^2(\RR^d)\cap L^{q_1}(\RR^d)$. 
Then Lemma~\ref{lemma-kernel} and Young's inequality imply 
$\widetilde{V}_{\theta}(\xi^2+1)^{-1}:L^2(\RR^d)\to L^2(\RR^d)$. 
Once we prove this estimate, we see that $\widetilde{V}_{\theta}(\xi^2+1)^{-1}:L^2(\RR^d)\to L^2(\RR^d)$ is analytic 
by estimating $(\widetilde{V}_{\theta}-\widetilde{V}_{\theta'})(\xi^2+1)^{-1}(\theta-\theta')^{-1}$ 
by the same arguments. 
We also see $\widetilde{V}_{\theta}(\xi^2+1)^{-1}$ is approximated in operator norm by operators 
with smooth integral kernels which vanish for $|\xi-\eta|\gg 1$ and for $|\xi-\eta|\ll 1$. 
By approximating $(\xi^2+1)^{-1}$ by compactly supported functions, we conclude that 
$\widetilde{V}_{\theta}(\xi^2+1)^{-1}$ is approximated in operator norm by operators 
with compactly supported smooth integral kernels, which are compact. 
\end{proof}

The discrete eigenvalues of $H_{\theta}$ for complex $\theta$ are resonances of $H$. 
The definition of resonances are also stated without referring to $H_{\theta}$. 
Resonances are poles of meromorphic continuation of $(u_1, R_+(z)u_2)$ for 
$u_1, u_2\in L^2_{\mathrm{comp}}(\RR^d)$, where $R_+(z)=(z-H)^{-1}$ for $\IM z>0$. 
The multiplicity of resonance $z$ is defined as the maximal number $k$ such that there exist 
$u_{1, 1}, \dots, u_{1, k}$ and $u_{2, 1}, \dots, u_{2, k}$ in $L^2_{\mathrm{comp}}(\RR^d)$ such that 
the matrix 
\[
\biggl(\frac{1}{2\pi i}\int_{|\zeta-z|=\ve} (u_{1, j}, R_{+}(\zeta)u_{2, \ell})d\zeta\biggr)_{j, \ell=1}^k
\] 
for $0<\ve\ll 1$ is invertible. 
We set 
\[
\Omega_{z, \ve}=(\RE z-\ve, \RE z+\ve)+i(\IM z-\ve, \infty)
\]
for small $\ve>0$. 
If we only consider the meromorphic continuation of $(u_1, R_+(z)u_2)$ in regions of the form $\Omega_{z, \ve}$, 
resonances are defined without ambiguity. 
If $\sigma_{\mathrm{ess}}(H_{\theta}) \cap \Omega_{z, \ve}=\emptyset$, then 
$\mathrm{Res}(H)=\sigma_d (H_{\theta})$ near $z$ including multiplicities. 
We used the fact that $\{U_{\theta}\hat{u}|\, u\in L^2_{\mathrm{comp}}(\RR^d)\}$ is dense in $L^2(\RR^d)$ 
for $\theta$ in $\{\theta\in \CC|\, (|\RE \theta|+|\IM \theta|)\mathrm{Lip}(v)<1\}$, 
which follows from the proof of \cite[Theorem~3]{H}. 
We take the vector field $v(\xi)$ such that $\xi \cdot v(\xi)>0$ near the energy level which we are interested in. 
For instance, we can consider $\theta=i\delta$ with $-\min\{\delta_0 E_0^{-1/2}, c_0\}<\delta<0$ 
by taking a compactly supported $v(\xi)$ such that $v(\xi)=\xi$ for $|\xi|<E_0^{1/2}$. 
Then $T(\Phi_{\theta}(\xi))=(1-\delta^2)\xi^2+2i\delta \xi^2$ for $|\xi|<E_0^{1/2}$. 
Thus resonances are defined in $\Omega(\delta_0, c_0)$. 


\subsection{Fourier distortion for discrete \Sch operators}\label{subsec-2}
Let $V_h: h\ZZ^d\to \RR$ be a potential function, and let 
$v_h\in \Cinf(h^{-1}\TT^d; \RR^d)$ be a vector field on $h^{-1}\TT^d$, 
where $\TT=\RR/2\pi \ZZ$. 
We set 
\[
\Phi_{h, \theta}(\xi)=\xi+\theta v_h(\xi), \quad U_{h, \theta}f_h(\xi)=J_{h, \theta}(\xi)^{1/2}f_h(\Phi_{h, \theta}(\xi)), 
\quad \xi\in h^{-1}\TT, 
\]
where $J_{h, \theta}(\xi)=\det \Phi_{h, \theta}'(\xi)$. 
We denote the discrete Fourier transform (or the Fourier series expansion) 
$\mathcal{F}_h: \ell^2(h\ZZ^d) \to L^2(h^{-1}\TT^d)$ by 
\begin{align*}
\mathcal{F}_h u_h(\xi)=(2\pi)^{-d/2}h^d \sum_{n\in \ZZ^d} u_h(hn)e^{-ihn\cdot \xi} 
\end{align*}
and set 
\[
\widetilde{H}_h=\mathcal{F}_h H_h \mathcal{F}_h^{-1}=\widetilde{T}_h(\xi)+\widetilde{V}_h, 
\]
where $\widetilde{T}_h(\xi)=2h^{-2}\sum_{j=1}^d(1-\cos h\xi_j)$ and 
$\widetilde{V}_h$ is the convolution operator on $h^{-1}\TT^d$ by $(2\pi h^{-1})^{-d/2}\mathcal{F}_h V_h$. 
We let 
\[
\widetilde{H}_{h, \theta}=U_{h, \theta} \widetilde{H}_h U_{h, \theta}^{-1}
=\widetilde{T}_h(\Phi_{h, \theta}(\xi))+\widetilde{V}_{h, \theta}, 
\]
where $\widetilde{V}_{h, \theta}$ has the integral kernel 
\[
\widetilde{V}_{h, \theta}(\xi, \eta)=(2\pi h^{-1})^{-d/2}J_{h, \theta}(\xi)^{1/2}(\mathcal{F}_h V_h)
(\Phi_{h, \theta}(\xi)-\Phi_{h, \theta}(\eta))J_{h, \theta}(\eta)^{1/2}. 
\] 
We set 
\[
H_{h, \theta}=\mathcal{F}_h^{-1} \widetilde{H}_{h, \theta} \mathcal{F}_h,\quad 
T_{h, \theta}=\mathcal{F}_h^{-1} \widetilde{T}_h(\Phi_{h, \theta}(\xi)) \mathcal{F}_h, \quad 
V_{h, \theta}=\mathcal{F}_h^{-1} \widetilde{V}_{h, \theta} \mathcal{F}_h.
\]

\begin{asm}\label{asm-res}
There exist $0<c_1\le 1$, $\delta_1>0$ and $\mu>0$ such that 
$\mathcal{F}_h V_h$ is analytic in $\{\xi+i\eta \in \CC^d/(2\pi h^{-1}\ZZ^d)|\, 
|\eta|<\min\{\delta_1, c_1|\xi|\}\}$ and 
\[
|\mathcal{F}_h V_h(\xi+i\eta)|\lesssim |\xi|^{-d+\mu}
\]
in $\{\xi+i\eta \in \CC^d/(2\pi h^{-1}\ZZ^d)|\, 
|\eta|<\min\{\delta'_1, c'_1|\xi|\}\}$ for any $0<\delta'_1<\delta_1$ and $0<c'_1<c_1$. 
\end{asm}
Here $|\xi|$ is the distance between $\xi$ and $0$ in $h^{-1}\TT^d$. 
If $|V_h(x)|\lesssim e^{-\delta_1 |x|}$, then $V_h$ satisfies Assumption~\ref{asm-res} for this 
$\delta_1$ and $c_1=1$. 
Another example is the restriction to $h\ZZ^d$ of an exterior dilation analytic potential. 
In fact, if $V$ satisfies Assumption~\ref{asm-1}, then its restriction $V_h$ to $h\ZZ^d$ satisfies 
Assumption~\ref{asm-res} for $c_1=c_0$ and $\delta_1=\delta_0$ by the proof of Lemma~\ref{lemma-kernel}
and Poisson's formula  
\[
(\mathcal{F}_h V_h)(\xi)=h^{-d/2}\sum_{m\in \ZZ^d}\hat{V}(\xi+2\pi h^{-1}m). 
\]
In this case, we thus have 
\begin{equation}\label{periodic}
\widetilde{V}_{h, \theta}f_h(\xi)=
(2\pi)^{-d/2}\int_{\RR^d}J_{h, \theta}(\xi)^{1/2}\hat{V}
(\Phi_{h, \theta}(\xi)-\Phi_{h, \theta}(\eta))J_{h, \theta}(\eta)^{1/2}f_h(\eta)d\eta, 
\end{equation}
where $f_h(\xi)$ is regarded as a periodic function. 

\begin{prop}\label{prop-2}
Under Assumption~\ref{asm-res}, 
$H_{h, \theta}$ is an analytic family of bounded operators. 
The spectrum of 
$H_{h, \theta}$ is discrete in the unbounded component of 
$\CC \setminus \{\widetilde{T}_h(\Phi_{h, \theta}(\xi))|\, \xi\in h^{-1}\TT^d\}$. 
Here $\theta$ ranges over 
\[
\bigl\{\theta\in \CC \bigm|\, \mathrm{Lip}(v_h)(c_1|\RE\theta|+|\IM\theta|)<c_1, \|v_h\|_{\infty}|\IM \theta|<\delta_1\bigr\}.
\] 
\end{prop}
\begin{proof}
As in the proof of Lemma~\ref{lemma-kernel}, the integral kernel $\widetilde{V}_{h, \theta}(\xi, \eta)$ of 
$\widetilde{V}_{h, \theta}$ is bounded by $\Psi_h(\xi-\eta)$ with $\Psi_h\in L^1(h^{-1}\TT^d)$.  
Then we see that $\widetilde{V}_{h, \theta}$ is an analytic family of compact operators 
since it is approximated by operators with smooth integral kernels in operator norm by Young's inequality. 
The Fredholm theory implies the discreteness of the spectrum.
\end{proof}
The discrete eigenvalues of $H_{h, \theta}$ are resonances of $H_h$ 
and the set of resonances is denoted by $\mathrm{Res}(H_h)$.
We choose the vector field $v_h$ so that $v_h(\xi)\cdot \nabla \widetilde{T}_h(\xi)>0$ near the 
energy surface we are interested in. 
A typical example is $v_h(\xi)=(\sin h\xi_1, \dots, \sin h\xi_d)$, 
Then resonances are defined in a complex neighborhood of 
$[0, 4dh^{-2}]\setminus\{0, 4h^{-2}, 8h^{-2}, \dots, 4dh^{-2}\}$. 
In the proof of Theorem~\ref{thm-1}, we use a different $v_h$. 

We next discuss the analytic vectors of the distortion on the torus. 
We set 
\[
\widehat{\mathcal{A}}_h=\bigl\{f\in \Cinf(h^{-1}\TT^d)\bigm|\, f \,\,\text{is analytic in}\,\, \CC^d/(2\pi h^{-1}\ZZ^d)\bigr\}.
\] 
Then $U_{h, \theta}f$ is analytic with respect to $\theta$ for $f \in \widehat{\mathcal{A}}_h$. 
\begin{lemma}
If $\mathrm{Lip}(v_h)(|\RE\theta|+|\IM \theta|)<1$ 
then $\{U_{h, \theta}f|\, f \in \widehat{\mathcal{A}}_h\}$ is dense in 
$L^2(h^{-1}\TT^d)$. 
\end{lemma}
\begin{proof}
Take $g\in \Cinf_c((-\pi h^{-1}, \pi h^{-1})^d)$ and we approximate it. 
As in \cite[Theorem~3]{H}, if we set 
\[
g_{1, k}(\xi)=\left(\frac{k}{\pi}\right)^{d/2}\int_{\RR^d}J_{h, \theta}(\eta)g(\eta)e^{-k(\xi-\Phi_{h, \theta}(\eta))^2}d\eta, 
\]
then $g_{1, k}$ is an entire analytic function and 
$\lim_{k\to \infty}g_{1, k}(\Phi_{h, \theta}(\bcdot))=g$ strongly in $L^2(\RR^d)$. 
Note that  $g_{1, k}(\xi)$ is exponentially small both for $k$ and $|\xi|^2$ outside $(-\pi h^{-1}, \pi h^{-1})^d$. 
Thus if we set $g_{2, k}(\xi)=\sum_{m \in \ZZ^d}g_{1, k}(\xi+2h^{-1}\pi m)$, 
then $g_{2, k}\in \widehat{\mathcal{A}}_h$ and 
$\lim_{k\to \infty}g_{2, k}(\Phi_{h, \theta}(\bcdot))=g$ strongly in $L^2(h^{-1}\TT^d)$. 
Since $J_{h, \theta}(\xi)^{\pm 1}$ is bounded, we complete the proof. 
\end{proof}

Then by the identity 
\[
(U_{h, \theta}f_1, (\widetilde{H}_{h, \theta}-z)^{-1} U_{h, \theta}f_2)=
(f_1, (\widetilde{H}_{h}-z)^{-1} f_2) 
\]
for $f_1, f_2\in \widehat{\mathcal{A}}_h$, the resonances are characterized as poles of meromorphic continuation of 
$(f_1, (\widetilde{H}_{h}-z)^{-1} f_2)$ for $f_1, f_2\in \widehat{\mathcal{A}}_h$. 
Note that $\mathcal{A}_h:=\mathcal{F}_h^{-1}\widehat{\mathcal{A}}_h$ 
is the space of super exponentially decaying functions on $h\ZZ^d$. 
We also note that any $f\in \widehat{\mathcal{A}}_h$ is approximated by $\mathcal{F}_h u$ for compactly supported 
$u\in \ell^2(h\ZZ^d)$ locally uniformly in $\CC^d/(2\pi h^{-1}\ZZ^d)$. 
Thus we conclude that resonances are characterized in terms of the original 
operator $H_h$ rather than the complex distorted operator $H_{h, \theta}$ as follows.  
Set $R_{h, +}(z)=(z-H_{h})^{-1}$ for $\IM z>0$. 
Resonances are the poles of meromorphic continuation of 
$(u_1, R_{h, +}(z)u_2)$ for compactly supported $u_1, u_2\in \ell^2(h\ZZ^d)$. 
The multiplicity of resonance $z$ is defined as the maximal number $k$ such that there exist 
compactly supported 
$u_{1, 1}, \dots, u_{1, k}$ and $u_{2, 1}, \dots, u_{2, k}$ in $\ell^2(h\ZZ^d)$ such that 
the matrix $\left(\frac{1}{2\pi i}\int_{|\zeta-z|=\ve} (u_{1, j}, R_{h, +}(\zeta)u_{2, \ell})d\zeta\right)_{j, \ell=1}^k$ 
for $0<\ve\ll 1$ is invertible. 
The definition is equivalent even if we replace compactly supported functions in $\ell^2(h\ZZ^d)$ with 
functions in $\mathcal{A}_h$. 
Moreover, resonances coincide with the discrete eigenvalues of $H_{h, \theta}$ including multiplicities. 
More precisely, we consider the meromorphic continuation of $(u_1, R_{h, +}(z)u_2)$ 
in the region of the form $\Omega_{z, \ve}=(\RE z-\ve, \RE z+\ve)+i(\IM z-\ve, \infty)$ for small $\ve>0$ 
when we discuss the resonances near $z$. Then resonances are defined without ambiguity. 
If $\sigma_{\mathrm{ess}}(H_{h, \theta}) \cap \Omega_{z, \ve}=\emptyset$, then 
$\mathrm{Res}(H_h)=\sigma_d (H_{h, \theta})$ near $z$ including multiplicities.

\begin{remark}
In the case of $v_h(\xi)=(\sin h\xi_1, \dots, \sin h\xi_d)$, it may be more natural to assume that 
$\mathcal{F}_h V_h$ has an analytic continuation to the region    
\[
\bigl\{\xi+i\eta \in \CC^d/(2\pi h^{-1}\ZZ^d)|\, 
\bigm|\eta_j|<\min\{\delta_1, c_1|\xi_j|\}\,\, \text{for any}\,\, j\bigr\}
\]
with the analogous bound 
as in Assumption~\ref{asm-res}. 
This is satisfied if we assume $|V_h(x)|\lesssim e^{-\delta_1 (|x_1|+\dots+|x_d|)}$. 
Then $\widetilde{H}_{h, \theta}$ is well-defined for complex $\theta$ in a neighborhood of 
$\{\theta=i\delta\in i\RR \mid  h|\delta|<c_1, |\delta|<\delta_1\}$.  
\end{remark}


\section{Continuum limit for continuous potentials}

\subsection{Abstract scheme}

Let $\mathcal{H}$, $\mathcal{H}_h$, $0<h<1$, be Hilbert spaces. 
In this subsection, all objects with the index $h$ are defined for $0<h<1$ unless otherwise stated. 
Let $I_h:\mathcal{H}_h\to \mathcal{H}$ be an isometry, which implies that 
$I_h^*:\mathcal{H}\to \mathcal{H}_h$ is a surjective partial isometry. 
Let $T$ and $T_h$ are closed operators on $\mathcal{H}$ and $\mathcal{H}_h$ respectively. 
Let $V$ and $V_h$ are bounded operators on $\mathcal{H}$ and $\mathcal{H}_h$ respectively. 
These are not assumed to be self-adjoint. Set $H=T+V$ and $H_h=T_h+V_h$, which are not self-adjoint in general. 
Recall that $\rho(T)$ is the resolvent set of $T$. 
\begin{lemma}\label{lemma-0}
Suppose that there exists $z_0\in \rho(T)$ such that $z_0\in \rho(T_h)$ for small $h>0$ and
\[
\lim_{h \to 0}\|I_h(T_h-z_0)^{-1}I_h^*-(T-z_0)^{-1}\|=0. 
\]
Suppose moreover that 
\[
\lim_{h \to 0}\|I_h V_h-V I_h\|=0.
\] 
Then for any $z_1\in \rho(H)$, there exist $h_0>0$ and $\gamma>0$ such that 
$z\in \rho(H_h)$ for $0<h<h_0$ and $|z-z_1|<\gamma$, and 
\[
\lim_{h \to 0}\|I_h(H_h-z)^{-1}I_h^*-(H-z)^{-1}\|=0
\]
uniformly for $|z-z_1|<\gamma$. 
In particular, $\lim_{h\to 0}\sigma_d(H_h)=\sigma_d(H)$ including algebraic multiplicities in 
$\CC \setminus \sigma_{\mathrm{ess}}(H)$. 
\end{lemma}
The last statement means that for any $z\in \CC \setminus \sigma_{\mathrm{ess}}(H)$ and 
small $\gamma>0$, there exists $h_0>0$ such that for $0<h<h_0$, 
the spectrum of $H_h$ is discrete in $\{\zeta\in \CC|\, |\zeta-z|<\gamma\}$  
and the number of eigenvalues of $H_h$ in $\{\zeta\in \CC|\, |\zeta-z|<\gamma\}$ counted with their algebraic 
multiplicities is equal to the algebraic multiplicity of $z$ as an eigenvalue of $H$, which is zero if $z \in \rho(H)$. 
This type of result is well-known for the self-adjoint case when $\mathcal{H}=\mathcal{H}_h$, $V=V_h=0$ and 
$I_h$ is the identity operator (see~\cite[Section~VII.\,7]{RS}).

\begin{proof}[Proof of Lemma~\ref{lemma-0}]
For simplicity, we denote the identity operator on $\mathcal{H}$ 
or $\mathcal{H}_h$ by $1$. 
For small $\gamma>0$, 
\[
H-z=(1+(V+z_0-z)(T-z_0)^{-1})(T-z_0)
\]
is invertible for $|z-z_1|<\gamma$. 
We have
\begin{align*}
1+(V+z_0-z)(T-z_0)^{-1}&=1+(V+z_0-z)I_h(T_h-z_0)^{-1}I_h^*+o(1)\\
&=1+I_h(V_h+z_0-z)(T_h-z_0)^{-1}I_h^*+o(1)
\end{align*}
by our assumptions. In this proof, $o(1)$ is with respect to the operator norm when $h\to 0$. 
Since this is invertible and $1+I_h(V_h+z_0-z)(T_h-z_0)^{-1}I_h^*$ is diagonal 
with respect to the decomposition $\mathcal{H}=\mathrm{Ran}I_h \bigoplus (\mathrm{Ran}I_h)^{\bot}$, 
we see that $1+(V_h+z_0-z)(T_h-z_0)^{-1}$ is invertible for small $h>0$ and  
\begin{align*}
&(H-z)^{-1}\\
&=(T-z_0)^{-1}(1+(V+z_0-z)(T-z_0)^{-1})^{-1}\\
&=(I_h(T_h-z_0)^{-1}I_h^*+o(1))((1+I_h(V_h+z_0-z)(T_h-z_0)^{-1}I_h^*)^{-1}+o(1))\\
&=I_h(T_h-z_0)^{-1}(1+(V_h+z_0-z)(T_h-z_0)^{-1})^{-1}I_h^*+o(1)\\
&=I_h(H_h-z_0)^{-1}I_h^*+o(1), 
\end{align*}
which completes the proof. 
\end{proof}


\subsection{Embedding operator}
In this subsection, we recall the embedding operator used in \cite{NT}.  
We set $\mathcal{H}=L^2 (\RR^d)$ and $\mathcal{H}_h=\ell^2(h\ZZ^d)$. 
In this paper, the norm of $\ell^2(h\ZZ^d)$ is defined as $\|u_h\|^2=h^d \sum_{n\in \ZZ^d}|u(hn)|^2$ for 
$u_h \in \ell^2(h\ZZ^d)$. 
We fix $\vp\in \mathscr{S}(\RR^d)$ and define $I_h:\mathcal{H}_h\to \mathcal{H}$ by 
\[
I_hu_h(x)=\sum_{n\in \ZZ^d}\vp(h^{-1}(x-nh))u_h(hn) 
\]
for $u_h\in \ell^2(h\ZZ^d)$. 
Then 
\[
I_h^* u(nh)=h^{-d}\int_{\RR^d}\overline{\vp(h^{-1}(x-nh))}u(x)dx 
\]
for $u\in L^2 (\RR^d)$. We assume that 
\[
(2\pi)^d\sum_{m\in \ZZ^d}|\hat{\vp}(\xi+2\pi m)|^2=1, 
\]
which is equivalent to the condition that $I_h$ is an isometry. 
Moreover, we assume that $\supp \hat{\vp}\subset (-2\pi, 2\pi)^d$. 

Recall that $\TT=\RR/2\pi\ZZ$. 
Set $\widehat{\mathcal{H}}=L^2(\RR^d)$ and $\widehat{\mathcal{H}}_h=L^2(h^{-1}\TT^d)$. 
Then $\mathcal{F}: \mathcal{H} \to \widehat{\mathcal{H}}$ and 
$\mathcal{F}_h: \mathcal{H}_h \to \widehat{\mathcal{H}}_h$ are unitary. 
We set $\widetilde{I}_h=\mathcal{F}_h I_h \mathcal{F}_h^*$. 
Then
\[
\widetilde{I}_hf_h(\xi)=(2\pi)^{d/2}\hat{\vp}(h\xi)f_h(\xi) 
\]
for $f_h\in \widehat{\mathcal{H}}_h$, where $f_h$ in the right hand side 
is regarded as a periodic function. 
Moreover, 
\[
{\widetilde{I}_h}^{^{\,*}} f(\xi)=(2\pi)^{d/2}\sum_{m \in \ZZ^d}\overline{\hat{\vp}(h\xi+2\pi m)}f(\xi+2\pi m h^{-1}) 
\]
for $f\in \widehat{\mathcal{H}}$.


\subsection{The kinetic term}\label{subsec-kinetic}
In the rest of this section, we assume Assumption~\ref{asm-1} and 
adopt the notation in Subsection~\ref{subsec-1.1}. 
We fix any $\Omega \Subset \Omega(\delta_0, c_0)$ and consider resonances in $\Omega$. 
We fix a compactly supported vector field $v(\xi)$ and $\theta$ such that 
$\mathrm{Res}(H)=\sigma_d(H_{\theta})$ in $\Omega$  
as discussed at the end of Subsection~2.1.  
We set $v_h(\xi)=\sum_{m\in \ZZ^d}v(\xi+2\pi m h^{-1})$, which defines a vector field on $h^{-1}\TT^d$. 
We take $h_0>0$ such that $\supp v \subset (-\pi h_0^{-1}, \pi h_0^{-1})^d$ and only consider $0<h<h_0$. 
Then $v_h$ is just the periodic extension of $v$. 
Since $\lim_{h\to 0}\widetilde{T}_h(\Phi_{h, \theta}(\xi))=\widetilde{T}(\Phi_{\theta}(\xi))$ 
locally uniformly with respect to $\xi$, 
we see that $\mathrm{Res}(H_h)=\sigma_d(H_{h, \theta})$ in $\Omega$ for small $h>0$. 
Thus it is enough to apply Lemma~\ref{lemma-0} to $H_{\theta}=T_{\theta}+V_{\theta}$ 
and $H_{\theta,h}=T_{h, \theta}+V_{h, \theta}$ to prove Theorem~\ref{thm-1}. 

We first discuss the kinetic term. 
It is easy to see that  
$(\widetilde{T}(\Phi_{\theta}(\xi))-z_0)^{-1}$ 
and $(\widetilde{T}_h(\Phi_{h, \theta}(\xi))-z_0)^{-1}$ 
are bounded uniformly for small $h>0$ if $-\RE z_0\gg 1$.  
We fix such $z_0$. 
\begin{lemma}\label{lemma-1}
\[
\|I_h(T_{h, \theta}-z_0)^{-1}I_h^*-(T_{\theta}-z_0)^{-1}\|
=\mathcal{O}(h^2) \,\,\,\text{when}\,\,\,h\to 0.  
\]
Hence, 
\[
\|(T_{h, \theta}-z_0)^{-1}I_h^*-I_h^*(T_{\theta}-z_0)^{-1}\|=\mathcal{O}(h^2) 
\] 
and
\[
\|(1-I_h I_h^*)(T_{\theta}-z_0)^{-1}\|=\mathcal{O}(h^2). 
\]
\end{lemma}
 
\begin{proof}
The proof is similar to \cite[Lemma~2.2, Lemma~2.3]{NT}, 
where the last two statements with $\theta=0$ were proved. 
Note that the last two statements follow from the first one since $I_h^* I_h=1$ and $(1-I_h I_h^*) I_h=0$.
For $f\in \widehat{\mathcal{H}}$, 
$\widetilde{I}_h(\widetilde{T}_h(\Phi_{h, \theta}(\xi))-z_0)^{-1}{\widetilde{I}_h}^{^{\,*}}  f$ 
is given by 
\begin{align*}
(2\pi)^d\hat{\vp}(h\xi)
(\widetilde{T}_h(\Phi_{h, \theta}(\xi))-z_0)^{-1}\sum_{m \in \ZZ^d}\overline{\hat{\vp}(h\xi+2\pi m))}f(\xi+2\pi m h^{-1}). 
\end{align*}
By the support property of $\hat{\vp}$, $\hat{\vp}(h\xi)\overline{\hat{\vp}(h\xi+2\pi m))}\not\equiv 0$ 
only for $m\in \{0, \pm 1\}^d$ and thus the sum is in fact a finite sum.
If $m\not =0$ and $\hat{\vp}(h\xi)\overline{\hat{\vp}(h\xi+2\pi m))}\not =0$, we see 
$|\widetilde{T}_h(\Phi_{h, \theta}(\xi))|\gtrsim h^{-2}$ for small $h>0$. Thus the contribution from $m\not=0$ terms is 
bounded by $Ch^2 \|f\|$. 
Thus it is enough to estimate the multiplication operator by 
\begin{align*}
(2\pi)^d |\hat{\vp}(h\xi)|^2
(\widetilde{T}_h(\Phi_{h, \theta}(\xi))-z_0)^{-1}-(\widetilde{T}(\Phi_{\theta}(\xi))-z_0)^{-1}.
\end{align*} 
If $(2\pi)^d|\hat{\vp}(h\xi)|^2\not=1$, the first term is bounded by $Ch^2$ since 
$h\xi \in (-2\pi+c, 2\pi-c)^d \setminus (-c, c)^d$ 
if moreover $|\hat{\vp}(h\xi)|^2\not=0$. 
The second term is also bounded by $Ch^2$ since $h|\xi|>c$. 
If $(2\pi)^d|\hat{\vp}(h\xi)|^2=1$, then $v_h(\xi)=v(\xi)$ and we have to estimate 
\begin{align*}
(\widetilde{T}_h(\Phi_{\theta}(\xi))-z_0)^{-1}(\widetilde{T}_h(\Phi_{\theta}(\xi))-\widetilde{T}(\Phi_{\theta}(\xi)))
(\widetilde{T}(\Phi_{\theta}(\xi))-z_0)^{-1} 
\end{align*}
and we claim that this is bounded by $Ch^2$. 
By Taylor's theorem, we see 
\begin{align*}
|\widetilde{T}_h(\Phi_{\theta}(\xi))-\widetilde{T}(\Phi_{\theta}(\xi))|\le Ch^{-2}|h\Phi_{\theta}(\xi)|^4\le Ch^2(1+|\xi|^4).  
\end{align*}
(If we use $v(0)=0$, the bound is $Ch^2|\xi|^4$.) If $(2\pi)^d|\hat{\vp}(h\xi)|^2=1$, then we see  
$|\widetilde{T}_h(\Phi_{\theta}(\xi))|\gtrsim |\xi|^2$. Thus we have the desired estimate. 
 \end{proof}


\subsection{The potential term}

\begin{lemma}\label{lemma-2}
Under Assumption~\ref{asm-1}, $\lim_{h\to 0}\|I_h V_{h, \theta}-V_{\theta} I_h\|=0$. 
\end{lemma}

\begin{proof}
By the equation~(\ref{periodic}) in Subsection~\ref{subsec-2}, 
it is enough to estimate the $L^2$-norm on $\RR^d$ of 
\begin{equation}\label{eq-3}
\begin{split}
&\hat{\vp}(h\xi)(2\pi)^{-d/2}\int_{\RR^d} J_{h, \theta}(\xi)^{1/2}\hat{V}(\Phi_{h, \theta}(\xi)-\Phi_{h, \theta}(\eta)) 
J_{h, \theta}(\eta)^{1/2}f_h(\eta)d\eta\\
&-(2\pi)^{-d/2}\int_{\RR^d} J_{\theta}(\xi)^{1/2}\hat{V}(\Phi_{\theta}(\xi)-\Phi_{\theta}(\eta)) 
J_{\theta}(\eta)^{1/2}\hat{\vp}(h\eta)f_h(\eta)d\eta 
\end{split}
\end{equation}
for $f_h\in \widehat{\mathcal{H}}_h$, which is seen as a periodic function. 
We use Lemma~\ref{lemma-kernel} and its extension to $\hat{V}(\Phi_{h, \theta}(\xi)-\Phi_{h, \theta}(\eta))$ below. 
We first show that we only have to consider $|\eta-\xi|<C$ for $C\gg 1$. 
This is immediate for the second term of (\ref{eq-3}) by the Schur's lemma and 
\[
\|\hat{\vp}(h\eta)f_h(\eta)\|_{L^2_{\eta}(\RR^d)}\le 2^d\|f_h\|_{\widehat{\mathcal{H}}_h}.
\] 
The integral kernel of the first term of (\ref{eq-3}) 
as an operator $L^2((-h^{-1}\pi, h^{-1}\pi)^d)\to L^2(\RR^d)$ is given by 
\begin{align*}
(2\pi)^{-d/2}\sum_{m\in \ZZ^d}
\hat{\vp}(h\xi)J_{h, \theta}(\xi)^{1/2}
\hat{V}(\Phi_{h, \theta}(\xi)-\Phi_{h, \theta}(\eta+2\pi mh^{-1}))J_{h, \theta}(\eta)^{1/2}. 
\end{align*}
If we apply the Schur's lemma to each term for $|\eta+2\pi mh^{-1}-\xi|>C$ 
and sum up the estimates, we see that the first term of (\ref{eq-3}) for $|\eta-\xi|>C$ is also small. 

If $\hat{\vp}(h\xi)\not=0$ or $\hat{\vp}(h\eta)\not=0$ in addition to $|\eta-\xi|<C$, then for small $h>0$, 
$\Phi_{h, \theta}(\xi)=\Phi_{\theta}(\xi)$, $\Phi_{h, \theta}(\eta)=\Phi_{\theta}(\eta)$, 
$J_{h, \theta}(\xi)=J_{\theta}(\xi)$ and $J_{h, \theta}(\eta)=J_{\theta}(\eta)$ since 
$v_h=v$ on $(-2h^{-1}\pi+C_1, 2h^{-1}\pi-C_1)^d$ for some $C_1>0$. 
Thus it is enough to estimate 
\begin{align*}
\int_{|\eta-\xi|<C} J_{\theta}(\xi)^{1/2}\hat{V}(\Phi_{\theta}(\xi)-\Phi_{\theta}(\eta)) 
J_{\theta}(\eta)^{1/2}(\hat{\vp}(h\xi)-\hat{\vp}(h\eta))f_h(\eta)d\eta. 
\end{align*}
By the support property of $\hat{\vp}$ and the condition that $|\eta-\xi|<C$, we only have to consider 
$\eta \in (-2h^{-1}\pi, 2h^{-1}\pi)^d$. Since $|(\hat{\vp}(h\xi)-\hat{\vp}(h\eta))|\lesssim h$, the Schur's lemma 
implies that 
\begin{align*}
&\left\|\int_{|\eta-\xi|<C} J_{\theta}(\xi)^{1/2}\hat{V}(\Phi_{\theta}(\xi)-\Phi_{\theta}(\eta)) 
J_{\theta}(\eta)^{1/2}(\hat{\vp}(h\xi)-\hat{\vp}(h\eta))f_h(\eta)d\eta\right\|_{L^2_{\xi}(\RR^d)}\\
&\le Ch\|f_h\|_{L^2((-2h^{-1}\pi, 2h^{-1}\pi)^d)}=2^d Ch\|f_h\|_{\widehat{\mathcal{H}}_h}, 
\end{align*}
which completes the proof. 
\end{proof}
Lemma~\ref{lemma-0}, Lemma~\ref{lemma-1} and Lemma~\ref{lemma-2} imply 
\[
\lim_{h\to 0}\|I_h(H_{h, \theta}-z)^{-1}I_h^*-(H_{\theta}-z)^{-1}\|=0
\]
for $z\in \rho(H_{\theta})$ and complete the proof of Theorem~\ref{thm-1}. 

\begin{remark}
If we assume $|\hat{V}(\xi+i\eta)|\lesssim |\xi|^{-\sigma}$ for $|\xi|>1$ with $\sigma>d$ in Assumption~\ref{asm-1}, 
we have 
\[
\|I_h V_{h, \theta}-V_{\theta} I_h\|=\mathcal{O}(h^{(\sigma-d)/(\sigma-d+1)})
\]
by considering $|\xi-\eta|\gtrless h^{-1/(\sigma-d+1)}$ in the proof of Lemma~\ref{lemma-2}. Thus 
\[
\|I_h(H_{h, \theta}-z)^{-1}I_h^*-(H_{\theta}-z)^{-1}\|=\mathcal{O}(h^{(\sigma-d)/(\sigma-d+1)})
\]
when $h\to 0$ in this case. 
\end{remark}


\section{Continuum limit for singular potentials}
In this section, we assume Assumption~\ref{asm-2} and 
adopt the notation in Subsection~\ref{subsec-1.2}. 
We note that Poisson's formula implies
\begin{equation}\label{Poisson}
(\mathcal{F}_h V_h)(\xi)=(2\pi)^{d/2}h^{-d/2}\sum_{m\in \ZZ^d}\hat{\phi}(h\xi+2\pi m)\hat{V}(\xi+2\pi h^{-1}m). 
\end{equation}
We need to estimate the convolution operator by $h^{d/2}\mathcal{F}_h V_h$ 
and its complex deformation. For that, we use the following simple lemma. 
\begin{lemma}\label{lemma-sum}
For $\psi\in \mathscr{S}(\RR^d)$, there exists $0\le C_{\psi}<\infty$ such that 
\[
\biggl\| \sum_{m\in \ZZ^d} \psi(h\xi+2\pi m) u(\xi+2\pi m h^{-1}) \biggr\|_{L_{\xi}^p([-h^{-1}\pi, h^{-1}\pi]^d)}
\le C_{\psi}\|u\|_{L^p(\RR^d)}
\]
for any $h>0$, $1\le p \le \infty$ and $u\in L^p(\RR^d)$. 
\end{lemma}
\begin{proof}
The left hand side is bounded by 
\[
\sum_{m \in \ZZ^d}\|\psi(\xi+2\pi m)\|_{L_{\xi}^{\infty}([-\pi, \pi]^d)} \|u\|_{L^p(\RR^d)}.
\]
Since $\psi\in \mathscr{S}(\RR^d)$, we see that 
$C_{\psi}:=\sum_{m \in \ZZ^d}\|\psi(\xi+2\pi m)\|_{L_{\xi}^{\infty}([-\pi, \pi]^d)}<\infty$. 
\end{proof}
For instance, we see 
\[
\sup_{0<h<1}\|h^{d/2}\mathcal{F}_h V_h\|_{L^1(h^{-1}\TT^d)+L^p(h^{-1}\TT^d)}<\infty
\]
if $\hat{V}\in L^1(\RR^d)+L^p(\RR^d)$. 
The complex distortion is defined as in the beginning of Subsection~\ref{subsec-kinetic}. 
\begin{lemma}\label{lemma-4}
Under Assumption~\ref{asm-2}, $\lim_{h\to 0}\|(T-i)^{-1}(I_h V_{h, \theta}-V_{\theta} I_h)\|=0$. 
\end{lemma}

\begin{proof}
We set $V^h=V*\phi_h$, which implies $V_h=V^h|_{h\ZZ^d}$. 
Let $p$ be as in Assumption~\ref{asm-2} and $p^{-1}+q^{-1}=1$. 
Then $q=2$ for $d=1, 2, 3$ and $q>d/2$ for $d\ge 4$. 
we take $p_1>1$ with $q^{-1}+p_1^{-1}=1/2$. 
In particular, $p_1=\infty$ for $d=1, 2, 3$. 
Since $(\widetilde{T}(\xi)-i)^{-1}\in L^{\infty}(\RR^d)\cap L^q(\RR^d)$, H\"{o}lder's inequality implies that 
it is enough to estimate the norm in $L^2(\RR^d)+L^{p_1}(\RR^d)$ of 
\begin{equation}\label{eq-4}
\begin{split}
&\hat{\vp}(h\xi)(2\pi)^{-d/2}\int_{\RR^d} J_{h, \theta}(\xi)^{1/2}\hat{V}^h(\Phi_{h, \theta}(\xi)-\Phi_{h, \theta}(\eta)) 
J_{h, \theta}(\eta)^{1/2}f_h(\eta)d\eta\\
&-(2\pi)^{-d/2}\int_{\RR^d} J_{\theta}(\xi)^{1/2}\hat{V}(\Phi_{\theta}(\xi)-\Phi_{\theta}(\eta)) 
J_{\theta}(\eta)^{1/2}\hat{\vp}(h\eta)f_h(\eta)d\eta 
\end{split}
\end{equation}
for $f_h\in \widehat{\mathcal{H}}_h$, which is seen as a periodic function. 

As in the proof of Lemma~\ref{lemma-kernel}, our assumption on $V$ implies that 
\[
|\hat{V}^h(\Phi_{h, \theta}(\xi)-\Phi_{h, \theta}(\eta))|\le \Psi(\xi-\eta)\psi(h\xi-h\eta), 
\]
where $\Psi \in L^1(\RR^d)+L^p(\RR^d)$ and $\psi\in \mathscr{S}(\RR^d)$. 
We show that the contribution from $|\eta-\xi|>C$, $C\gg 1$, is small by the following arguments. 
Note that $p^{-1}+1/2=p_1^{-1}+1$. 
Since $|\eta-\xi|>C$, $C\gg 1$, we may assume that $\|\Psi\|_{L^1+L^p}\ll 1$. 
The $L^2+L^{p_1}$-norm of the second term of (\ref{eq-4}) for $|\eta-\xi|>C$ is small by Young's inequality. 
The integral kernel of the first term of (\ref{eq-4}) as an operator 
$L^2((-h^{-1}\pi, h^{-1}\pi)^d)\to L^2(\RR^d)+L^{p_1}(\RR^d)$ is given by 
\begin{align*}
(2\pi)^{-d/2}\sum_{m\in \ZZ^d}
\hat{\vp}(h\xi)J_{h, \theta}(\xi)^{1/2}
\hat{V}^h(\Phi_{h, \theta}(\xi)-\Phi_{h, \theta}(\eta+2\pi mh^{-1}))J_{h, \theta}(\eta)^{1/2}. 
\end{align*}
This is bounded by  
\[
C_1\sum_{m\in \ZZ^d}\Psi(\xi-\eta-2\pi mh^{-1})\psi(h\xi-h\eta-2\pi m) 
\]
for some $C_1>0$. Since $\xi \in (-2\pi h^{-1}, 2\pi h^{-1})^d$, 
$2^d$-time applications of Young's inequality on $h^{-1}\TT^d$ imply that the first term of (\ref{eq-4}) for 
$|\eta-\xi|>C$ is small by Lemma~\ref{lemma-sum} and $\|\Psi\|_{L^1+L^p}\ll 1$.  

Then as in the proof of Lemma~\ref{lemma-2}, we only have to consider $|\eta-\xi|<C$ and 
$\eta \in (-2h^{-1}\pi, 2h^{-1}\pi)^d$. 
Then $\Phi_{h, \theta}=\Phi_{\theta}$ and $J_{h, \theta}=J_{\theta}$. 
We recall that $\hat{V}(\xi)-\hat{V}^h(\xi)=\hat{V}(\xi)(1-(2\pi)^{d/2}\hat{\phi}(h\xi))$ and 
$(2\pi)^{d/2}\hat{\phi}(0)=1$. 
Since $|\eta-\xi|<C$, 
\[
1-(2\pi)^{d/2}\hat{\phi}(h(\Phi_{\theta}(\xi)-\Phi_{\theta}(\eta)))=\mathcal{O}(h),
\]
and we may think
\[
|\hat{V}(\Phi_{\theta}(\xi)-\Phi_{\theta}(\eta))|\le \Psi_1(\xi-\eta)
\]
with $\Psi_1\in L^1$. Then Young's inequality implies
\begin{align*}
&\biggl\|\int_{|\eta-\xi|<C} J_{\theta}(\xi)^{1/2}(\hat{V}-\hat{V}^h)(\Phi_{\theta}(\xi)-\Phi_{\theta}(\eta)) 
J_{\theta}(\eta)^{1/2}\hat{\vp}(h\xi)f_h(\eta)d\eta\biggr\|_{L_{\xi}^2(\RR^d)}\\
&\qquad \le Ch\|f_h\|_{L^2((-2h^{-1}\pi, 2h^{-1}\pi)^d)}=2^d Ch\|f_h\|_{\widehat{\mathcal{H}}_h}. 
\end{align*}
The remaining term is 
\begin{align*}
&\int_{|\eta-\xi|<C} J_{\theta}(\xi)^{1/2}\hat{V}(\Phi_{\theta}(\xi)-\Phi_{\theta}(\eta)) 
J_{\theta}(\eta)^{1/2}(\hat{\vp}(h\xi)-\hat{\vp}(h\eta))f_h(\eta)d\eta 
\end{align*} 
and its $L^2$-norm is $\mathcal{O}(h)$ by exactly the same argument as in the proof of Lemma~\ref{lemma-2}, 
which completes the proof.
\end{proof}

\begin{lemma}\label{lemma-5}
Under Assumption~\ref{asm-2}, $V_{h, \theta}$ is infinitesimally $T_{h, \theta}$-bounded uniformly for $0<h<1$. 
\end{lemma}
\begin{proof}
The statement means that for any $\ve>0$ there exists $C_{\ve}>0$ such that 
\[
\|V_{h, \theta} u_h\|_{\ell^2(h\ZZ^d)}
\le \ve \|T_{h, \theta}u_h\|_{\ell^2(h\ZZ^d)}+C_{\ve}\|u_h\|_{\ell^2(h\ZZ^d)}
\]
for any $0<h<1$ and $u_h \in \ell^2(h\ZZ^d)$, which is equivalent to 
\[
\lim_{t\to \infty} \sup_{0<h<1}\|\widetilde{V}_{h, \theta}
(\widetilde{T}_h(\Phi_{h, \theta}(\xi))-it)^{-1}\|_{L^2(h^{-1}\TT^d)\to L^2(h^{-1}\TT^d)}=0. 
\]
As in the proof of Lemma~\ref{lemma-4}, we take $p$ from Assumption~\ref{asm-2} and 
set $p^{-1}+q^{-1}=1$. Recall that $q=2$ for $d=1, 2, 3$ and $q>d/2$ for $d\ge 4$. 
Then 
\[
\sup_{0<h<1}\|(\widetilde{T}_h(\Phi_{h, \theta}(\xi))-it)^{-1}\|_{L^{\infty}(h^{-1}\TT^d)\cap L^q(h^{-1}\TT^d)}
\]
is small for $t\gg 1$, which follows from the observation: 
\[
|\widetilde{T}_h(\Phi_{h, \theta}(\xi)) |\gtrsim |\xi|^2\quad  \text{for} 
\quad \xi \in (-\pi h^{-1}, \pi h^{-1})^d.
\]
Take $q_1$ with $q^{-1}+1/2=q_1^{-1}$. 
Then the operator norm of 
\[
(\widetilde{T}_h(\Phi_{h, \theta}(\xi))-it)^{-1}\ :\ L^2(h^{-1}\TT^d)\to L^2(h^{-1}\TT^d)\cap L^{q_1}(h^{-1}\TT^d)
\]
is small for $t\gg 1$ uniformly for $0<h<1$ by H\"{o}lder's inequality. 
Using the extension of Lemma~\ref{lemma-kernel} to $\hat{V}(\Phi_{h, \theta}(\xi)-\Phi_{h, \theta}(\eta))$,  
the equation~(\ref{Poisson}) and Lemma~\ref{lemma-sum}, 
we see that the integral kernel of $\widetilde{V}_{\theta, h}$ on $h^{-1}\TT^d$ satisfies 
$|\widetilde{V}_{\theta, h}(\xi, \eta)|\le \Psi_h(\xi-\eta)$,  
where $\sup_{0<h<1}\|\Psi_h\|_{L^1(h^{-1}\TT^d)+L^p(h^{-1}\TT^d)}<\infty$. 
Since $p^{-1}+q_1^{-1}=1/2+1$, Young's inequality completes the proof. 
\end{proof}

\begin{proof}[Proof of Theorem~\ref{thm-2}]
The proof is similar to that in Nakamura-Tadano~\cite[Subsection~2.3]{NT} using 
Lemma~\ref{lemma-4} and Lemma~\ref{lemma-5}. We briefly recall it.
By Lemma~\ref{lemma-0} (with $V=0$ there), it is enough to prove 
\[
\lim_{h \to 0}\|I_h(H_{h, \theta}-z_0)^{-1}I_h^*-(H_{\theta}-z_0)^{-1}\|=0 
\]
for some $z_0\in \rho(H_{\theta})$. 
We write 
\begin{align*}
&I_h(H_{h, \theta}-z_0)^{-1}I_h^*-(H_{\theta}-z_0)^{-1}\\
&=I_h(H_{h, \theta}-z_0)^{-1}(I_h^*H_{\theta}-H_{h, \theta}I_h^*)(H_{\theta}-z_0)^{-1}
-(1-I_h I_h^*)(H_{\theta}-z_0)^{-1}. 
\end{align*}
Since $V_{\theta}$ is infinitesimally $T_{\theta}$-bounded by the proof of Proposition~\ref{prop-2}, 
we see that $(T_{\theta}-z_0)(H_{\theta}-z_0)^{-1}$ is bounded. 
Thus Lemma~\ref{lemma-1} implies that the second term tends to zero when $h\to 0$. 
The first term is decomposed to 
\begin{equation}\label{eq-thm-2}
\begin{split}
&I_h(H_{h, \theta}-z_0)^{-1}(I_h^*V_{\theta}-V_{h, \theta}I_h^*)(H_{\theta}-z_0)^{-1}\\
&\quad +I_h(H_{h, \theta}-z_0)^{-1}(I_h^*T_{\theta}-T_{h, \theta}I_h^*)(H_{\theta}-z_0)^{-1}. 
\end{split}
\end{equation}
The first term of (\ref{eq-thm-2}) tends to zero when $h\to 0$ by the adjoint estimate of Lemma~\ref{lemma-4}. 
We note that Lemma~\ref{lemma-5} implies that 
$(H_{h, \theta}-z_0)^{-1}(T_{h, \theta}-z_0)$ is  
uniformly bounded for $0<h<1$. 
Thus the operator norm of the second term of (\ref{eq-thm-2}) is bounded by 
\begin{align*}
&C\|(T_{h, \theta}-z_0)^{-1}(I_h^*T_{\theta}-T_{h, \theta}I_h^*)(T_{\theta}-z_0)^{-1}\|\\
&=C\|(T_{h, \theta}-z_0)^{-1}I_h^*-I_h^*(T_{\theta}-z_0)^{-1}\|,  
\end{align*}
which tends to zero by Lemma~\ref{lemma-1}. 
\end{proof} 

\begin{proof}[Proof of Theorem~\ref{thm-3}]
Theorem~\ref{thm-3} is proved if we set $\theta=0$ in the proof of Theorem~\ref{thm-2}. 
\end{proof}

\begin{remark}
If we assume $|\hat{V}(\xi+i\eta)|\lesssim |\xi|^{-\sigma}$ for $|\xi|>1$ with $p\sigma>d$ in Assumption~\ref{asm-2}, 
we have 
\[
\|(T-i)^{-1}(I_h V_{h, \theta}-V_{\theta} I_h)\|=\mathcal{O}(h^{(p\sigma-d)/(p\sigma-d+p)})
\]
by considering $|\xi-\eta|\gtrless h^{-p/(p\sigma-d+p)}$ in the proof of Lemma~\ref{lemma-4}. Thus 
\[
\|I_h(H_{h, \theta}-z)^{-1}I_h^*-(H_{\theta}-z)^{-1}\|=\mathcal{O}(h^{(p\sigma-d)/(p\sigma-d+p)})
\]
for $z\in \rho(H_{\theta})$ when $h\to 0$ in this case. Similarly, we have 
\[
\|I_h(H_h-z)^{-1}I_h^*-(H-z)^{-1}\|=\mathcal{O}(h^{(p\sigma-d)/(p\sigma-d+p)})
\]
for $z\in \rho(H)$ if we assume $|\hat{V}(\xi)|\lesssim |\xi|^{-\sigma}$ for $|\xi|>1$ with $p\sigma>d$ in Theorem~\ref{thm-3}. 
When $d=1, 2, 3$, we assume $\sigma>d/2$ and the bound is $\mathcal{O}(h^{(2\sigma-d)/(2\sigma-d+2)})$ since we take $p=2$. 
When $d\ge 4$, we assume $\sigma>d-2$ and the bound is $\mathcal{O}(h^{(\sigma-d+2)/(\sigma-d+3)-\ve})$ for any $\ve>0$ since 
we take $p<d/(d-2)$.
For instance, we have the bound $\mathcal{O}(h^{1/3})$ in the Coulomb potential case since $d=3, \sigma=2$. 
\end{remark}

\section*{Acknowledgement}
KK is supported by JSPS Research Fellowship for Young Scientists, Grant Number JP23KJ2090.
SN is partially supported by JSPS Kakenhi Grant Number 21K03276. 


\end{document}